\documentclass[letterpaper,12pt,notoc]{JHEP3}
\def\pd{\partial}
\def\mc{\mathcal}

\usepackage{graphicx}
\usepackage{amsmath}
\usepackage{amssymb}

\preprint{ \hbox{}\hfill arXiv: 1210.8064}

\title{Holographic RG flows in six dimensional F(4) gauged
supergravity}
\author{Parinya Karndumri\\
String Theory and Supergravity Group, Department
of Physics, Faculty of Science, Chulalongkorn University, 254 Phayathai Road, Pathumwan, Bangkok 10330, Thailand\\
Thailand Center of Excellence in Physics, CHE, Ministry of Education, 319 Dusit, Bangkok 10400, Thailand \\
E-mail: \email{parinya.ka@hotmail.com}}

\abstract{We study critical points of $F(4)$ gauged supergravity in
six dimensions coupled to three vector multiplets. Scalar fields are
described by $\mathbb{R}^+\times \frac{SO(4,3)}{SO(4)\times SO(3)}$
coset space, and the gauge group is given by $SO(3)_R\times SO(3)$
with $SO(3)_R$ being the R-symmetry. The maximally supersymmetric
critical point with all scalars vanishing preserves the full
$SO(3)_R\times SO(3)$ symmetry. This is dual to a superconformal
field theories (SCFT$_5$) arising from a near horizon geometry of
the D4-D8 brane system in type I$'$ theory with an enhanced global
symmetry $E_1\sim SU(2)$. Apart from this trivial critical point, we
identify a new supersymmetric critical point preserving the full
supersymmetry with the $SO(3)_R\times SO(3)$ symmetry broken to its
diagonal subgroup. This critical point should correspond to a new
SCFT in five dimensions. We study an RG flow solution interpolating
between the SCFT with $E_1$ symmetry and the new supersymmetric
critical point. The flow describes a supersymmetric deformation
driven by a vacuum expectation value of relevant operators of
dimension $3$. We identify the dual operators with the mass terms
for hypermultiplet scalars in the dual field theory. The solution
provides an example of analytic supersymmetric RG flows in
AdS$_6$/CFT$_5$ correspondence.}

\begin{document}
\section{Introduction}
The study of holographic renormalization group (RG) flow is one of
the most important applications of the AdS/CFT correspondence
\cite{maldacena}. Soon after the original proposal of the
correspondence, many works considering RG flows in five dimensional
gauged supergravity have been done, see for example \cite{fgpw},
\cite{an} and \cite{gir}. These results describe various
perturbations of $N=4$ SYM in four dimensions. Since the AdS/CFT
correspondence has been extended to other dimensions not only for
the duality between an $AdS_5$ supergravity and a CFT$_4$
\cite{AdS3CFT2,AdS4CFT3,nishimura,AdS7CFT6,AdS9CFT8}, it is interesting to study holographic RG flows
in these extensions as well. Until now, a lot of works on
holographic RG flows in three and four dimensional gauged
supergravities have appeared, see for example \cite{bs}, \cite{gkn},
\cite{AP}, \cite{waner4D_flow} and \cite{membrane_flow}.
\\
\indent On the other hand, a study of AdS$_6$/CFT$_5$ correspondence has not been
explored in details although some works in this direction can be
found in \cite{nishimura} and \cite{ferrara_AdS6}, see also
\cite{Bergman} for a more recent result. To the best of the author's
knowledge, only one holographic RG flow solution, studied in
\cite{F4_nunezAdS6}, in AdS$_6$/CFT$_5$ correspondence has been
studied so far. This work will provide another example which is
related to a deformation of five dimensional $N=2$ SYM at a
conformal fixed point with enhanced global symmetry $SU(2)$.
\\
\indent The starting point is the $F(4)$ gauged supergravity in six
dimensions with $N=(1,1)$ supersymmetries. The pure $F(4)$ gauged
supergravity has been constructed in \cite{F4_Romans}. It has been
known since its first construction that the scalar potential of the
pure $F(4)$ gauged supergravity admits two critical points
\cite{F4_Romans}. One of them is maximally supersymmetric, and the
other one breaks all supersymmetries. The latter is however stable
and should correspond to a non-supersymmetric conformal field theory
according to the general principle of the AdS/CFT correspondence. A
numerical RG flow solution interpolating between these two critical
points has been given in \cite{F4_nunezAdS6}. This pure $F(4)$ gauged
supergravity has also been studied in the context of holography in
\cite{F4_nunez} in which RG flows between a UV CFT$_5$ and an IR
CFT$_3$ or CFT$_2$ have been given.
\\
\indent In this work, we consider an RG flow in AdS$_6$/CFT$_5$
correspondence from matter coupled $F(4)$ gauged supergravity
constructed in \cite{F4SUGRA1} and \cite{F4SUGRA2}. The dual CFT's
with $N=2$ supersymmetry are fixed points of the $N=2$ $USp(2k)$ SYM
theory describing the worldvolume theory of D4-branes
\cite{Seiberg_5Dfield}, see also \cite{Seiberg_5Dfield2}. The fixed
points corresponding to interacting CFT's arise in an infinite
coupling limit. In term of brane configurations, they can be
described as near horizon geometries of the D4-D8 brane system in
type I$'$ theory \cite{D4D8}. The configuration should also be
obtained in massive type IIA theory where the D8-branes are known to
exist \cite{ferrara_AdS6}. At the fixed points, the $SO(2N_f)\times U(1)$ global symmetry
gets enhanced to $E_{N_f+1}$, $N_f=0,\ldots, 7$ with $E_1=SU(2)$,
$E_2=SU(2)\times U(1)$, $E_3=SU(3)\times SU(2)$, $E_4=SU(5)$ and
$E_5=SO(10)$ \cite{Seiberg_5Dfield}. $E_{6,7,8}$ are the usual
exceptional groups. $N_f$ denotes the number of flavor
hypermultiplets or equivalently the number of D8-branes. The
extension to theories with global symmetry $\tilde{E}_1=U(1)$ and
$E_0=\mathbf{I}$ (no global symmetry) has been studied in
\cite{Seiberg_5Dfield3}. The matter coupled $F(4)$ gauged
supergravity whose scalars paramatrized by $\mathbb{R}^+\times
\frac{SO(4,n)}{SO(4)\times SO(n)}$ coset space can give rise to
gravity duals of these 5-dimensional CFTs with enhanced global
symmetry by gauging the $SU(2)_R\times G$ subgroup of $SO(4)\times
SO(n)$. $n$ denotes the number of matter multiplets, vector
multiplets in this case. $SU(2)_R$ is the R-symmetry given by the
diagonal subgroup of $SU(2)\times SU(2)\sim SO(4)$, and $G$ is
identified with the flavor symmetry $E_{N_f+1}$ with
$n=\textrm{dim}\, G$.
\\
\indent The full symmetry at the fixed points is given by
$F(4)\times G$. The $F(4)$ is the superconformal group containing
$SU(2)_R\times SO(5,2)$ as its bosonic subgroup. The correspondence
between $F(4)$ gauged supergravity and these five dimensional CFTs
has been discussed in \cite{ferrara_AdS6}, and some relations
between supergravity fields with the global symmetry $G$ being $E_7$
and their dual operators have been identified. These relations have
been verified to be consistent with the spectrum of the matter
coupled $F(4)$ gauged supergravity theory in \cite{F4SUGRA1}.
\\
\indent We will consider a simple case of the matter coupled $F(4)$
gauged supergravity in which there are three matter (vector)
multiplets. This case corresponds to $N_f=0$ or, equivalently, no
D8-branes present. We will study the scalar potential of the theory
and identify some of the critical points. This gives the first
non-trivial critical point in matter coupled $F(4)$ gauged
supergravity and describes a new supersymmetric fixed point of the
dual field theory in five dimensions. We then compute the full
scalar mass spectrum at the critical point which will give
information on dimensions of the dual operators. We will also study
an RG flow solution corresponding to a supersymmetric deformation of
the UV $N=2$ CFT with global $SU(2)$ symmetry by a vacuum
expectation value of dimension-3 operators.
\\
\indent The paper is organized as follow. In section \ref{F4_sugra},
we give some formulae in the matter coupled $F(4)$ gauged
supergravity which forms a framework of the whole paper. We will
give our notations and conventions which are slightly different from
\cite{F4SUGRA1} and \cite{F4SUGRA2}. In section
\ref{critical_points}, we study the scalar potential and identify
new critical points of the matter coupled $F(4)$ gauged
supergravity. We will also give scalar mass spectra at each critical
point. An RG flow solution connecting the trivial critical point
with maximal supersymmetry and a new supersymmetric critical point
with the same amount of supersymmetry will be given in section
\ref{RGflow}. Finally, we give our conclusions and comments in
section \ref{conclusion}.

\section{Matter coupled $F(4)$ gauged supergravity}\label{F4_sugra}
In this section, we review the construction and relevant formulae
for the matter coupled $F(4)$ gauged supergravity. Although we will
work with the metric signature $(-+++++)$, most of the notations and
conventions used in this work will be closely parallel to those
given in \cite{F4SUGRA1} and \cite{F4SUGRA2}. We refer the reader to
these references for a beautiful geometric construction. Some
gaugings of this theory can be obtained from a truncation of the
maximal gauged supergravity in six dimensions. Moreover, these
gaugings can be described in an $O(1,1)\times SO(4,n)$ covariant
form using the embedding tensor formalism \cite{Henning_6DN4}.
However, in this paper, we will restrict ourselves to the
construction of \cite{F4SUGRA1} and \cite{F4SUGRA2}.
\subsection{General
matter coupled $F(4)$ gauged supergravity} The $F(4)$ supergravity
is a half maximal, $N=(1,1)$, supergravity in six dimensions. The
field content of the supergravity multiplet is given by
\begin{displaymath}
\left(e^a_\mu,\psi^A_\mu, A^\alpha_\mu, B_{\mu\nu}, \chi^A,
\sigma\right)
\end{displaymath}
where $e^a_\mu$, $\chi^A$ and $\psi^A_\mu$ denote the graviton, the
spin $\frac{1}{2}$ field and the gravitino, respectively. Space-time and tangent space indices are denoted by $\mu,\nu,\ldots$ and $a,b,\ldots$.
Both$\chi^A$ and $\psi^A_\mu$ are the eight-component
pseudo-Majorana spinor with indices $A,B=1,2$ referring to the
fundamental representation of the $SU(2)_R$ R-symmetry. The
remaining fields are given by the dilaton $\sigma$, four vectors
$A^\alpha_\mu,\, \alpha=0,1,2,3$, and a two-form field
$B_{\mu\nu}$. As in \cite{F4SUGRA1}, it is more convenient in the
gauged theory to decompose the $\alpha$ index into $\alpha=(0,r)$ in
which $r=1,2,3$.
\\
\indent The matter field in the $N=(1,1)$ theory is given by the
vector multiplet whose field content is
\begin{displaymath}
(A_\mu,\lambda_A,\phi^\alpha).
\end{displaymath}
We will label each matter multiplet by an index $I=1,\ldots, n$. The
$4n$ scalars $\phi^{\alpha I}$ are described by a symmetric space
$SO(4,n)/SO(4)\times SO(n)$. Furthermore, the dilaton $\sigma$ can
be regarded as living in the coset space $\mathbb{R}^+\sim O(1,1)$
which is the duality group of the pure supergravity theory.
Together, the $4n+1$ scalars of the matter coupled theory live in
the coset space
\begin{equation}
\mathbb{R}^+\times \frac{SO(4,n)}{SO(4)\times SO(n)},
\end{equation}
and the global symmetry is given by $O(1,1)\times SO(4,n)$. The
$SU(2)_R$ R-symmetry is the diagonal subgroup of $SU(2)\times
SU(2)\sim SO(4)\subset SO(4)\times SO(n)$.
\\
\indent $O(1,1)$ can be parametrized by $e^\sigma$ while the
$SO(4,n)/SO(4)\times SO(n)$ coset will be parametrized by the coset
representative $L^\Lambda_{\phantom{as}\Sigma}$,
$\Lambda,\Sigma=0,\ldots , 3+n$. It is also convenient to split the
$L^\Lambda_{\phantom{as}\Sigma}$ into
$(L^\Lambda_{\phantom{as}\alpha},L^\Lambda_{\phantom{as}I})$ and
further to $(L^\Lambda_{\phantom{as}0}, L^\Lambda_{\phantom{as}r},
L^\Lambda_{\phantom{as}I})$. The composite connections $Q$ and the
vielbein of the coset $P$ are defined schematically via
\begin{equation}
L^{-1}dL=Q+P\, .
\end{equation}
\indent We now come to gaugings of the matter coupled theory and
will restrict ourselves to compact gaugings. The gauge group is a
subgroup of $SO(4)\times SO(n)\subset SO(4,n)$. In the pure gauged
supergravity, the gauge group is given by $SU(2)_R$. In the matter
coupled case, we consider the gauge group of the form $SU(2)_R\times
G$ with $\textrm{dim}\, G=n$. The restriction $\textrm{dim}\, G=n$
comes from the fact that there are only $n$ vector multiplets, or
equivalently the maximum number of the gauge fields for the gauge
group $G$ is given by $n$. We will denote the structure constant of
the full gauge group $SU(2)_R\times G$ by
$f^\Lambda_{\phantom{as}\Pi\Sigma}$ which can be split into
$\epsilon_{rst}$ and $C_{IJK}$ for $SU(2)_R$ and $G$, respectively.
\\
\indent As usual, the gauging is implemented by covariantizing all
the derivatives, and to restore the supersymmetry, fermionic
mass-like terms and a scalar potential are introduced. We only give
some needed information to our study of the vacua and RG flows and
refer the reader to \cite{F4SUGRA1} and \cite{F4SUGRA2} for more
details. Moreover, the supersymmetry transformations of fermions are
also modified by some shifts at first order in the gauge couplings.
The direct product structure of the gauge group $SU(2)_R\times G$
leads to two coupling constants $g_1$ and $g_2$. As has been shown
in \cite{F4SUGRA2}, without the mass parameter $m$ of the two-form
field the maximally supersymmetric $AdS_6$ vacuum does not exist.
Therefore, only the $m\neq 0$ case is relevant for the present work.
\\
\indent The bosonic Lagrangian under consideration here consists of
the metric and scalars \cite{F4SUGRA2}
\begin{equation}
\mathcal{L}=\frac{1}{4}eR-e\pd_\mu \sigma\pd^\mu \sigma
-\frac{1}{4}eP_{I\alpha\mu}P^{I\alpha\mu}-eV
\end{equation}
where $e=\sqrt{-g}$. The $\phi^{I\alpha}$ kinetic term is written in
term of $P^{I\alpha}_\mu=P^{I\alpha}_i\pd_\mu\phi^i$, $i=1,\ldots,
4n$. The scalar potential is given by \cite{F4SUGRA2}
\begin{eqnarray}
V&=&-e^{2\sigma}\left[\frac{1}{36}A^2+\frac{1}{4}B^iB_i+\frac{1}{4}\left(C^I_{\phantom{s}t}C_{It}+4D^I_{\phantom{s}t}D_{It}\right)\right]
+m^2e^{-6\sigma}\mc{N}_{00}\nonumber \\
& &-me^{-2\sigma}\left[\frac{2}{3}AL_{00}-2B^iL_{0i}\right]
\end{eqnarray}
where $\mc{N}_{00}$ is the $00$ component of the scalar matrix
\begin{equation}
\mc{N}_{\Lambda\Sigma}=L^{\phantom{as}0}_\Lambda
L^{-1}_{0\Sigma}+L^{\phantom{as}i}_\Lambda
L^{-1}_{i\Sigma}-L^{\phantom{as}I}_\Lambda L^{-1}_{I\Sigma}\, .
\end{equation}
Various quantities appearing in the scalar potential are defined as
follow
\begin{eqnarray}
A&=&\epsilon^{rst}K_{rst},\qquad B^i=\epsilon^{ijk}K_{jk0},\\
C^{\phantom{t}t}_I&=&\epsilon^{trs}K_{rIs},\qquad D_{It}=K_{0It}
\end{eqnarray}
where
\begin{eqnarray}
K_{rst}&=&g_1\epsilon_{lmn}L^l_{\phantom{r}r}(L^{-1})_s^{\phantom{s}m}L_{\phantom{s}t}^n+
g_2C_{IJK}L^I_{\phantom{r}r}(L^{-1})_s^{\phantom{s}J}L_{\phantom{s}t}^K,\nonumber
\\
K_{rs0}&=&g_1\epsilon_{lmn}L^l_{\phantom{r}r}(L^{-1})_s^{\phantom{s}m}L_{\phantom{s}0}^n+
g_2C_{IJK}L^I_{\phantom{r}r}(L^{-1})_s^{\phantom{s}J}L_{\phantom{s}0}^K,\nonumber
\\
K_{rIt}&=&g_1\epsilon_{lmn}L^l_{\phantom{r}r}(L^{-1})_I^{\phantom{s}m}L_{\phantom{s}t}^n+
g_2C_{IJK}L^I_{\phantom{r}r}(L^{-1})_I^{\phantom{s}J}L_{\phantom{s}t}^K,\nonumber
\\
K_{0It}&=&g_1\epsilon_{lmn}L^l_{\phantom{r}0}(L^{-1})_I^{\phantom{s}m}L_{\phantom{s}t}^n+
g_2C_{IJK}L^I_{\phantom{r}0}(L^{-1})_I^{\phantom{s}J}L_{\phantom{s}t}^K\,
.
\end{eqnarray}
\indent Other ingredients we are going to use are the supersymmetry
transformations of $\chi^A$, $\lambda^I_A$ and $\psi^A_\mu$
\begin{eqnarray}
\delta\psi_{\mu
A}&=&D_\mu\epsilon_A-\frac{1}{24}\left(Ae^\sigma+6me^{-3\sigma}(L^{-1})_{00}\right)\epsilon_{AB}\gamma_\mu\epsilon^B\nonumber
\\
& &-\frac{1}{8}
\left(B_te^\sigma-2me^{-3\sigma}(L^{-1})_{t0}\right)\gamma^7\sigma^t_{AB}\gamma_\mu\epsilon^B,\label{delta_psi}\\
\delta\chi_A&=&\frac{1}{2}\gamma^\mu\pd_\mu\sigma\epsilon_{AB}\epsilon^B+\frac{1}{24}
\left[Ae^\sigma-18me^{-3\sigma}(L^{-1})_{00}\right]\epsilon_{AB}\epsilon^B\nonumber
\\
& &-\frac{1}{8}
\left[B_te^\sigma+6me^{-3\sigma}(L^{-1})_{t0}\right]\gamma^7\sigma^t_{AB}\epsilon^B\label{delta_chi}\\
\delta
\lambda^{I}_A&=&-P^I_{ri}\pd_\mu\phi^i\sigma^{r}_{AB}\gamma^\mu\epsilon^B-P^I_{0i}
\pd_\mu\phi^i\gamma^7\gamma^\mu\epsilon_{AB}\epsilon^B-\left(2i\gamma^7D^I_{\phantom{s}t}-C^I_{\phantom{s}t}\right)
e^\sigma\sigma^t_{AB}\epsilon^B \nonumber
\\
& &+2me^{-3\sigma}e^{-3\sigma}(L^{-1})^I_{\phantom{ss}0}
\gamma^7\epsilon_{AB}\epsilon^B\label{delta_lambda}
\end{eqnarray}
where $\sigma^{tC}_{\phantom{sd}B}$ are Pauli matrices, and $\sigma^t_{AB}=\sigma^{tC}_{\phantom{sd}B}\epsilon_{CA}$ with
$\epsilon_{AB}=-\epsilon_{BA}$. In the above equations, we have
given only terms involving the metric and scalar fields. The
space-time gamma matrices $\gamma^\mu$ satisfy
\begin{equation}
\{\gamma^a,\gamma^b\}=2\eta^{ab},\qquad
\eta^{ab}=\textrm{diag}(-1,1,1,1,1,1),
\end{equation}
and $\gamma^7=\gamma^0\gamma^1\gamma^2\gamma^3\gamma^4\gamma^5$ with
$(\gamma^7)^2=\mathbf{I}$.

\subsection{$F(4)$ gauged supergravity coupled to three vector
multiplets} In this subsection, we consider the $F(4)$ gauged
supergravity coupled to three matter multiplets. The gauge group is
given by $SU(2)_R\times SO(3)$ with structure constants
$\epsilon_{rst}$ and $\epsilon_{IJK}$.
\\
\indent We begin with the $SO(4,3)/SO(4)\times SO(3)$ coset. It is
convenient to parametrize the group generators by basis elements
\begin{equation}
(e^{xy})_{zw}=\delta_{xz}\delta_{yw}, \qquad w,x,y,z=1,\ldots, 7
\end{equation}
from which we find the following generators:
\begin{eqnarray}
SO(4)&:&\qquad
J^{\alpha\beta}=e^{\beta+1,\alpha+1}-e^{\alpha+1,\beta+1},\qquad \alpha,\beta=0,1,2,3,\nonumber \\
SU(2)_{R}&:&\qquad J^{rs}=e^{s+1,r+1}-e^{r+1,s+1},\qquad r,s=1,2,3,\nonumber \\
SO(3)&:&\qquad T^{IJ}=e^{J+4,I+4}-e^{I+4,J+4},\qquad I,J,K=1,2,3\, .
\end{eqnarray}
Non-compact generators are given by
\begin{equation}
Y^{\alpha I}=e^{\alpha+1,I+4}+e^{I+4,\alpha+1}\, .
\end{equation}
\indent Explicit calculations show that it is extremely difficult
(if possible) to compute the scalar potential on the full
12-dimensional scalar manifold $SO(4,3)/SO(4)\times SO(3)$. As in
other similar works, we will employ the method introduced in
\cite{warner}. The potential is computed on a particular submanifold
of $SO(4,3)/SO(4)\times SO(3)$ which is invariant under some
subgroup of the gauge group $SU(2)_R\times SO(3)\sim SO(3)_R\times
SO(3)$. From now on, we will use $SU(2)_R$ and $SO(3)_R$,
interchangeably, since the two terminologies are convenient in
different contexts.
\\
\indent We will study the scalar potential on scalar fields
invariant under $SO(3)_{\textrm{diag}}$, $SO(2)_{\textrm{diag}}$,
$SO(2)_R$ and $SO(2)$. Under $SO(3)_R\times SO(3)$, the 12 scalars
transform as $(\mathbf{1}+\mathbf{3},\mathbf{3})$. Under
$SO(3)_{\textrm{diag}}\subset (SO(3)_R\times
SO(3))_{\textrm{diag}}$, they transform as
\begin{equation}
(\mathbf{1}+\mathbf{3})\times
\mathbf{3}=\mathbf{1}+\mathbf{3}_{\textrm{Adj}}+\mathbf{3}+\mathbf{5}
\end{equation}
where the $\mathbf{3}_{\textrm{Adj}}$ denoting the adjoint
representation. For convenience, the subscript Adj is used to
distinguish $\mathbf{3}_{\textrm{Adj}}$ from the vector
representation $\mathbf{3}$. We immediately see that there is one
singlet under $SO(3)_{\textrm{diag}}$. It corresponds to the
following generator
\begin{equation}
Y_s=Y^{21}+Y^{32}+Y^{43}\, .\label{Ys}
\end{equation}
\indent Under $SO(2)_R\subset SO(3)_R$ with the embedding
$\mathbf{3}\rightarrow \mathbf{1}+\mathbf{2}$, the 12 scalars
transform as $3\times (\mathbf{1}+\mathbf{1}+\mathbf{2})$ containing
six singlets given by
\begin{eqnarray}
\bar{Y}_1&=&Y^{11},\qquad \bar{Y}_2=Y^{12},\qquad
\bar{Y}_3=Y^{13},\nonumber \\
\bar{Y}_4&=&Y^{21},\qquad \bar{Y}_5=Y^{22},\qquad \bar{Y}_6=Y^{23}\,
.\label{Y_bar}
\end{eqnarray}
Under $SO(2)\subset SO(3)$ with the same embedding, there are four
singlets in the decomposition of the 12 scalars under $SO(2)$,
$4\times (\mathbf{1+\mathbf{2}})$. The singlets are given by
\begin{equation}
\tilde{Y}_1=Y^{13},\qquad \tilde{Y}_2=Y^{23},\qquad
\tilde{Y}_3=Y^{33},\qquad \tilde{Y}_4=Y^{43}\, .\label{Y_tilde}
\end{equation}
\indent Finally, under $SO(2)_{\textrm{diag}}\subset (SO(2)_R\times
SO(2))_{\textrm{diag}}$, the 12 scalars transform as
\begin{equation}
(\mathbf{1}+\mathbf{1}+\mathbf{2})\times
(\mathbf{1}+\mathbf{2})=\mathbf{1}+\mathbf{1}+\mathbf{1}+\mathbf{2}+\mathbf{2}+\mathbf{2}+\mathbf{3}
\end{equation}
in which there are three singlets corresponding to
\begin{equation}
\hat{Y}_1=Y^{13},\qquad \hat{Y}_2=Y^{23},\qquad
\hat{Y}_3=Y^{31}+Y^{42}\, .\label{Y_hat}
\end{equation}
\indent In subsequent sections, we will study the scalar potential
on these submanifolds together with the associated critical points
and possible RG flows.

\section{Critical points of the matter coupled $F(4)$ gauged
supergravity}\label{critical_points} In this section, we study some
critical points of the $F(4)$ gauged supergravity coupled to three
vector multiplets. We first compute the scalar potential of the
matter coupled theory and identify some critical points. We begin
with the study of the scalar potential on the submanifold invariant
under $SO(3)_{\textrm{diag}}\subset SO(3)_R\times SO(3)$. The coset
representative is given by
\begin{equation}
L=e^{b Y_s}\label{SO3_L}
\end{equation}
where $Y_s$ is given in \eqref{Ys}. We find the potential
\begin{eqnarray}
V&=&\frac{1}{16} e^{-6 a} \left[e^{8 a} \left[(\cosh (6 b)-9 \cosh
(2 b)) \left(g_1^2+g_2^2\right)-8 g_1 g_2 \sinh ^3(2 b)\right.\right.\nonumber \\
& &\left.\left.+8 (g_2^2-g_1^2) \right]-64 e^{4 a} m \left(g_1 \cosh
^3b-g_2 \sinh ^3b\right)+16 m^2\right].
\end{eqnarray}
where $a$ is the dilaton $\sigma$.
\\
\indent As shown in \cite{F4SUGRA1} and \cite{F4SUGRA2}, the trivial
critical point, given by setting $a=0$ and $b=0$, is maximally
supersymmetric provided that $g_1=3m$ with $g_2$ arbitrary. We can
study the scalar mass spectrum of the 13 scalars at this critical
point by expanding the potential on the full 13-dimensional scalar
manifold to quadratic order. This gives the following mass spectrum:
\begin{center}
\begin{tabular}{|c|c|}
  \hline
  scalars & $m^2L^2$ \\ \hline
  $(\mathbf{1},\mathbf{1})$ & $-6$ \\
  $(\mathbf{1},\mathbf{3})$ & $-4$ \\
  $(\mathbf{3},\mathbf{3})$ & $-6$ \\
  \hline
\end{tabular}
\end{center}
\indent We have labeled the scalars according to their
representations under $SO(3)_R\times SO(3)$ gauge symmetry of the
critical point. The AdS radius is given by
$L=\sqrt{-\frac{5}{V_0}}=\frac{1}{2m}$ with $V_0=-20m^2$. This
result exactly agrees with the analysis of \cite{F4SUGRA1} and
\cite{F4SUGRA2}. Using the discussion given in \cite{ferrara_AdS6},
the dilaton $(\mathbf{1},\mathbf{1})$ and the
$(\mathbf{3},\mathbf{3})$ scalars do not correspond to the highest
component in the supermultiplet and describe non-supersymmetric
deformations. On the other hand, the $(\mathbf{1},\mathbf{3})$
scalars gives rise to supersymmetric deformations since they
correspond to the highest component.
\\
\indent We then move to the first non trivial critical point. This
is given by
\begin{eqnarray}
a&=&\frac{1}{4}\ln\left[-\frac{3m\sqrt{g_2^2-g_1^2}}{g_1g_2}\right],\qquad
b=\frac{1}{2}\ln \left[\frac{g_2+g_1}{g_2-g_1}\right],\nonumber \\
V_0&=&-20m^2\left[-\frac{g_1g_2}{3m\sqrt{g_2^2-g_1^2}}\right]^{\frac{3}{2}},
\qquad
L=\frac{1}{2m}\left[-\frac{3m\sqrt{g_2^2-g_1^2}}{g_1g_2}\right]^{\frac{3}{4}}.
\end{eqnarray}
This critical point is supersymmetric as can be checked by using the
supersymmetry transformations given in \eqref{delta_psi},
\eqref{delta_chi} and \eqref{delta_lambda}. In some details, the
choice $b=\frac{1}{2}\ln \left[\frac{g_2+g_1}{g_2-g_1}\right]$ or
$b=0$ automatically gives $\delta\lambda^I_A=0$. The condition from
$\delta\chi_A=0$ determines the value of
$a=\frac{1}{4}\ln\left[-\frac{3m\sqrt{g_2^2-g_1^2}}{g_1g_2}\right]$.
Note also that the critical point is valid for $g_2<-g_1$ when
$g_1>0$. For $g_1<0$, we need to take $g_2<g_1$.
\\
\indent It is useful to compute scalar mass spectrum at this
critical point. This will give some useful information about the
dimensions of the dual operators in the holographic context since
this critical point will involve in an RG flow solution studied in
the next section. It is convenient to label the 13 scalars according
to their representations under the unbroken gauge symmetry
$SO(3)_{\textrm{diag}}$. We find the following scalar masses:
\begin{center}
\begin{tabular}{|c|c|}
  \hline
  scalars & $m^2L^2$ \\ \hline
  $\mathbf{1}$ & $-6$ \\
  $\mathbf{1}$ & $24$ \\
  $\mathbf{3}$ & $14$ \\
  $\mathbf{3}_{\textrm{Adj}}$ & $0$ \\
  $\mathbf{5}$ & $6$ \\
  \hline
\end{tabular}
\end{center}
\indent The first singlet in the table corresponds to the dilaton
while the second one is the singlet parametrized by our coset
representative \eqref{SO3_L}. There are precisely three massless
scalars corresponding to three Goldstone bosons of the symmetry
breaking $SO(3)_R\times SO(3)\rightarrow SO(3)_{\textrm{diag}}$.
These should give rise to three massive vector fields whose masses
we have not computed.
\\
\indent As in the pure $F(4)$ gauged supergravity case, there is
another critical point in which $a=-\frac{1}{4}\ln \frac{g_1}{m}$
and $b=0$. But, this critical point is unstable whenever the scalars
in the matter multiplets are turned on. The corresponding scalar
mass spectrum is given as follow:
\begin{center}
\begin{tabular}{|c|c|}
  \hline
  scalars & $m^2L^2$ \\ \hline
  $(\mathbf{1},\mathbf{1})$ & $10$ \\
  $(\mathbf{1},\mathbf{3})$ & $0$ \\
  $(\mathbf{3},\mathbf{3})$ & $-10$ \\
  \hline
\end{tabular}
\end{center}
In this table, the scalars are classified according their
representations under $SO(3)_R\times SO(3)$. It can be seen that the
$(\mathbf{3},\mathbf{3})$ scalars have mass squares below the BF
bound, $-\frac{25}{4}$. Therefore, this critical point is unstable.
\\
\indent There is another critical point which can be given,
numerically. We simply give its position on the scalar manifold
without going into the details here. At this critical point, we find
\begin{eqnarray}
a&=&-\frac{1}{4} \ln \left[\frac{3 \coth b}{16 \cosh (2 b)-6 \cosh
(4 b)+38}
\left[7 \sinh (3 b)+5 \sinh (5 b)\right.\right. \nonumber \\
& &\left.\left.+2 \sinh b \left(4 \sqrt{2} \cosh ^2b \sqrt{7 \cosh
(4 b)-5}-7\right)\right]\right],\label{a}
\end{eqnarray}
and $b$ is given implicitly in term of $g_2$ by
\begin{equation}
g_2=-\frac{3 m \left[3 \cosh (4 b)+4 \sqrt{2} \sqrt{7 \cosh (4
b)-5}+13\right] \coth b}{8 \cosh (2 b)-3 \cosh (4 b)+19}\, .
\end{equation}
From \eqref{a}, $a$ is real for $-0.82395<b<0.82395$. We will not
give the explicit expression for the cosmological constant here but
only make a comment that the critical point is $AdS_6$ for
$-0.64795<b<0.64795$. This critical point is however
non-supersymmetric. The analysis shows that all values of $b$ in
this range give rise to unstable $AdS_6$ critical points. As an
example, we give the full mass spectrum for a particular value of
$b=-\frac{1}{2}$ which gives
\begin{equation}
g_2=-0.5963m,\qquad a=0.1618,\qquad V_0=-22.8963m^2\, .
\end{equation}
The scalar masses are given in the table below.
\begin{center}
\begin{tabular}{|c|c|}
  \hline
  scalars & $m^2L^2$ \\ \hline
  $\mathbf{1}$ & $16.3581$ \\
  $\mathbf{1}$ & $-11.1726$ \\
  $\mathbf{3}$ & $-4$ \\
  $\mathbf{3}_{\textrm{Adj}}$ & $0$ \\
  $\mathbf{5}$ & $-2.1305$ \\
  \hline
\end{tabular}
\end{center}
From the table, we see that one of the singlet scalar, a combination
of the dilaton $a$ and the scalar $b$, has mass below the BF bound.
\\
\indent We have also studied the scalar potential on other scalar
submanifolds invariant under various $SO(2)$ residual gauge
symmetries. With the $SO(2)_{\textrm{diag}}$ invariant scalars, the
coset representative takes the form
\begin{equation}
L=e^{a_1\hat{Y}_1}e^{a_2\hat{Y}_2}e^{a_3\hat{Y}_3}\, .
\end{equation}
We find the scalar potential
\begin{eqnarray}
V&=&\frac{1}{4}  \left[e^{2 \sigma } \left[\sinh ^2(2 a_3)
\left(g_2^2 \cosh (2 a_1)+g_1^2 \cosh (2 a_2)\right)-4 g_1^2 \cosh
^2a_3-4 g_2^2 \sinh ^2a_3\right.\right.\nonumber
\\
& &\left.+8 g_2 \sinh ^2a_3  \cosh a_1 \sinh a_2 \cosh ^2a_3 (g_2
\cosh a_1 \sinh a_2-2 g_1 \cosh
a_2)\right]\nonumber \\
& & -16 m e^{-2 \sigma } \left(g_1 \cosh a_1 \cosh a_2 \cosh
^2a_3-g_2 \sinh a_2 \sinh^2a_3\right)\nonumber \\
& &\left.+4m^2e^{-6 \sigma }\cosh ^2a_1+4 m^2 e^{-6 \sigma }\sinh ^2a_1 \cosh (2 a_2)\right].
\end{eqnarray}
\indent For $SO(2)$ invariant scalars, the coset representative is
given by
\begin{equation}
L=e^{a_1\tilde{Y}_1}e^{a_2\tilde{Y}_2}e^{a_3\tilde{Y}_3}e^{a_4\tilde{Y}_4}
\end{equation}
giving rise to the following potential
\begin{eqnarray}
V&=&\frac{1}{4} m^2 e^{-6 \sigma} \left[4 \sinh^2a_1 \cosh ^2a_2
\left[\cosh ^2 a_3 \cosh (2 a_4)+\sinh ^2a_3\right]+2 \sinh ^2a_1
\cosh (2 a_2)\right.\nonumber \\
& &\left.+\cosh (2 a_1)+3\right] -4 g_1 m e^{-2 \sigma } \cosh a_1
\cosh a_2 \cosh a_3 \cosh a_4-g_1^2 e^{2 \sigma}.
\end{eqnarray}
\indent Finally, the coset representative for $SO(2)_R$ invariance
is parametrized by
\begin{equation}
L=e^{a_1\bar{Y}_1}e^{a_2\bar{Y}_2}e^{a_3\bar{Y}_3}e^{a_4\bar{Y}_4}e^{a_5\bar{Y}_5}e^{a_6\bar{Y}_6},
\end{equation}
and the corresponding potential is given by
\begin{eqnarray}
V&=&\frac{1}{8} e^{2 \sigma } \left[8 g_2^2 \cosh ^2a_1 \cosh ^2a_4
\left(\cosh a_2 \sinh a_3 \sinh a_5 \cosh a_6 -\sinh a_2 \sinh
a_6\right)^2\right.
\nonumber \\
& &+8 g_2^2 \left[\sinh a_1 \cosh a_5 \sinh a_6-\cosh a_1 \sinh a_4
\left(\cosh a_2 \sinh a_3 \cosh a_6
\right.\right.\nonumber \\
& &\left.\left.+\sinh a_2 \sinh a_5 \sinh a_6\right)\right]^2+8
g_2^2 \left(\cosh a_1 \sinh a_2 \sinh a_4 \cosh a_5 -\sinh a_1 \sinh
a_5\right)^2 \nonumber
\\
& &\left. + g_1^2 \left(4 \cosh ^2a_4 \cosh ^2a_5 \cosh (2 a_6)+2
\cosh (2 a_4)\cosh^2a_5+ \cosh (2 a_5)-7\right)\right]
\nonumber \\
& & -4 g_1 m e^{-2 \sigma } \cosh a_1 \cosh a_2 \cosh a_3 \cosh a_4
\cosh a_5 \cosh a_6 - g_1^2e^{2\sigma} \cosh ^2a_4 \cosh ^2a_5
\times
\nonumber \\
& &\cosh ^2a_6+m^2 e^{-6 \sigma } \left[\left(\cosh a_1 (\cosh a_2
\sinh a_3 \sinh a_6+\sinh a_2 \sinh a_5 \cosh a_6)
\right.\right.\nonumber \\
& &\left.+\sinh a_1 \sinh a_4 \cosh a_5 \cosh
a_6\right)^2+\left(\cosh a_1 \left(\cosh a_2 \sinh a_3 \cosh a_6
\right.\right.\nonumber \\
& &\left.\left.+\sinh a_2 \sinh a_5 \sinh a_6\right)+\sinh a_1 \sinh
a_4 \cosh a_5 \sinh a_6\right)^2 +\sinh ^2a_1 \cosh ^2a_4
\nonumber \\
& &\left.+\cosh ^2a_1 \cosh ^2a_2 \cosh ^2a_3+(\cosh a_1 \sinh a_2 \cosh a_5+\sinh a_1 \sinh a_4 \sinh a_5)^2\right].\nonumber \\
& &\label{poten}
\end{eqnarray}
\indent This potential can be further truncated to the $SO(3)_R$
invariant scalars by setting $a_4=a_5=a_6=0$. The resulting
potential is then given by
\begin{eqnarray}
V&=&\frac{1}{4} e^{-6 \sigma } \left[-16 g_1 m e^{4 \sigma } \cosh
a_1 \cosh a_2 \cosh a_3+m^2 \left[2 \cosh ^2a_1\right.\right.\times
\nonumber
\\ & & \left.\left.\left(2 \cosh ^2a_2 \cosh (2 a_3)+\cosh (2 a_2)\right)+\cosh (2 a_1)-3\right]-4 g_1^2 e^{8 \sigma }\right].\label{poten1}
\end{eqnarray}
This potential is interesting in the sense that if we set
$\sigma=0$, it involves only the $SO(3)_R$ singlet scalars which
correspond to supersymmetric deformations \cite{ferrara_AdS6}.
\\
\indent However, we have not found any interesting critical points
of \eqref{poten1} whether supersymmetric or not. Similarly, apart
from $L=\mathbf{I}_{7\times 7}$ and $\sigma=0$, we are not able to
figure out any critical points of the other $SO(2)$ invariant
potentials given above. The analysis shows that, most probably, they
might not admit non-trivial critical points at all, but a more
detailed analysis is needed particularly for a very complicated
potential in \eqref{poten}.

\section{RG flow solutions}\label{RGflow}
In this section, we will find an RG flow solution interpolating
between the trivial critical point and the $SO(3)_{\textrm{diag}}$
critical point. This will describe a supersymmetric deformation of
the UV $SO(3)_R\times SO(3)$ SCFT. We will also give a numerical RG
flow solution which describes a non-supersymmetric deformation of
the $N=2$ UV CFT with global symmetry $SU(2)$. However, this
solution in pure $F(4)$ gauged supergravity has already been studied
in \cite{F4_nunezAdS6}. Similar non-supersymmetric solutions in
four, five and seven dimensional gauged supergravities have been
given in \cite{non_SUSY7Dflow}, \cite{non_SUSYflow1} and
\cite{non_SUSYflow2}.

\subsection{A supersymmetric RG flow}
In order to study holographic RG flows, we make a standard domain
wall ansatz for the metric
\begin{equation}
ds^2=e^{2A(r)}dx^2_{1,4}+dr^2\label{DW_ansatz}
\end{equation}
where $dx^2_{1,4}$ is the metric on five-dimensional Minkowski
space. In the present case, we are mainly interested in
supersymmetric solutions. Therefore, we set up the BPS equations
from supersymmetry transformations of fermions given in
\eqref{delta_psi}, \eqref{delta_chi} and \eqref{delta_lambda} with
all but the metric and $SO(3)_{\textrm{diag}}$ singlet scalars $a$
and $b$ vanishing. Furthermore, Poincare symmetry in five dimensions
requires that both $a$ and $b$ can only be a function of the radial
coordinate $r$.
\\
\indent By imposing the projection condition
$\gamma_{\hat{r}}\epsilon_A=\epsilon_A$, we find the resulting BPS
equations
\begin{eqnarray}
b'&=&-\frac{1}{4}e^{a-3b}(e^{4b}-1)\left[(1+e^{2b})g_1+(1-e^{2b})g_2\right],\label{6D_eq1}\\
a'&=&-\frac{1}{16}e^{a-3b}\left[(1+e^{2b})^3g_1+(1-e^{2b})^3g_2\right]
+\frac{3}{2}me^{-3a},\label{6D_eq2}\\
A'&=&\frac{1}{16}e^{a-3b}\left[(1+e^{2b})^3g_1+(1-e^{2b})^3g_2\right]
+\frac{1}{2}me^{-3a}\label{6D_eq3}
\end{eqnarray}
where the $r$-derivative is denoted by $'$. From these equations, it
is clearly seen that there are two supersymmetric critical points as
previously identified.
\\
\indent To find an RG flow solution interpolating between these two
critical points, we need to solve the above BPS equations for
$b(r)$, $a(r)$ and $A(r)$. By defining a new radial coordinate
$\tilde{r}$ via $\frac{d\tilde{r}}{dr}=e^{a-3b}$, we can solve
equation \eqref{6D_eq1} for $b(\tilde{r})$. The solution for $b(\tilde{r})$
is given implicitly by
\begin{eqnarray}
\tilde{r}&=&\frac{4}{g_1+g_2}b-\frac{1}{2g_1}\ln(1-e^{2b})-\frac{1}{2g_2}\ln(1+e^{2b})\nonumber
\\
&
&+\frac{(g_1-g_2)^2}{g_1g_2(g_1+g_2)}\ln\left[(1+e^{2b})g_1+(1-e^{2b})g_2\right].
\end{eqnarray}
In this solution, we have neglected an additive integration constant
which can be removed by shifting the coordinate $\tilde{r}$.
\\
\indent Combining equations \eqref{6D_eq1} and \eqref{6D_eq2}, we
find
\begin{equation}
\frac{da}{db}=-\frac{e^{3b-4a}\left[24m-e^{4a-3b}
[(1+e^{2b})^3g_1+(1-e^{2b})^3g_2]\right]}
{4(e^{4b}-1)\left[(1+e^{2b})g_1+(1-e^{2b})g_2\right]}
\end{equation}
which can be solved to give a solution for $a$
\begin{equation}
a=\frac{1}{4}\ln\left[\frac{e^{-b}(6m+C_1(e^{4b}-1))}{(1+e^{2b})g_1+(1-e^{2b})g_2}\right].
\end{equation}
It can be easily shown that $a=0$ as $b=0$ for $g_1=3m$. In order to make this
solution interpolate between the trivial critical point and the new
critical points, we need to choose the constant $C_1$ to be
\begin{equation}
C_1=-\frac{3m(g_1-g_2)^2}{2g_1g_2}
\end{equation}
which finally leads to the solution for $a$
\begin{equation}
a=\frac{1}{4}\ln
\left[\frac{3me^{-b}[(1-e^{2b})g_1+(1+e^{2b})g_2]}{2g_1g_2}\right].
\end{equation}
The solution clearly goes to the IR fixed point with
$a=\frac{1}{4}\ln\left[-\frac{3m\sqrt{g_2^2-g_1^2}}{g_1g_2}\right]$
for $b=\frac{1}{2}\ln \left[\frac{g_2+g_1}{g_2-g_1}\right]$.
\\
\indent By the same procedure, we can combine equation
\eqref{6D_eq1} and \eqref{6D_eq3} to obtain
\begin{equation}
\frac{dA}{db}=-\frac{3g_2(1-e^{2b})^3+g_1(3+13e^{2b}+13e^{4b}+3e^{6b})}{12(e^{2b}-1)(1+e^{2b})
[(1+e^{2b})g_1+(1-e^{2b})g_2]}\, .
\end{equation}
We then find the solution for $A(r)$ up to an additive
integration constant that can be absorbed by rescaling the
coordinates in $dx^2_{1,4}$
\begin{equation}
A=\frac{1}{4}b-\frac{1}{3}\ln
(1-e^{2b})-\frac{1}{4}\ln(1+e^{2b})+\frac{1}{3}\ln[(1+e^{2b})g_1+(1-e^{2b})g_2].
\end{equation}
Along the flow, the solution preserves half of the supersymmetry, but the supersymmetry is enhanced at the two fixed points at which the $\gamma_{\hat{r}}$ projector is not imposed.
\\
\indent Near the UV point, the scalar fields behave as
\begin{equation}
a(r)\sim e^{-\frac{3r}{L_{\textrm{UV}}}} \qquad \textrm{and}\qquad b(r)\sim
e^{-\frac{3r}{L_{\textrm{UV}}}}
\end{equation}
where the $AdS_6$ radius in the UV is given by
$L_{\textrm{UV}}=\frac{1}{2m}$. By the relation
$m^2L^2=\Delta(\Delta-5)$, this corresponds to an RG flow driven by
a vacuum expectation value of relevant operator of dimension $3$,
see the discussion in \cite{Freedman_lecture}. The possibility for
$\Delta=2$ is excluded since there is no bosonic operator of
dimension $2$ in five dimensional field theories. According to the
relation between six dimensional supergravity fields and operators
in the dual five dimensional field theory \cite{ferrara_AdS6}, we
find that the dual operators to the scalars $a$ and $b$ should be
the scalar mass terms for the hypermultiplet scalars.
\\
\indent The linearized equations near the IR point give
\begin{equation}
a(r)\sim e^{-\frac{3r}{L_{\textrm{IR}}}},\qquad b(r)\sim
e^{\frac{3r}{L_{\textrm{IR}}}},\qquad
L_{\textrm{IR}}=\frac{1}{2m}\left(-\frac{3m\sqrt{g_2^2-g_1^2}}{g_1g_2}\right)^{\frac{3}{4}}\,
.
\end{equation}
We can see that the operator dual to $b(r)$ acquires an anomalous
dimension resulting in an irrelevant operator of dimension $8$ which
is in agreement with the value of its mass-square $m^2L^2=24$. This
is however not the case for $a(r)$ which is still dual to the
operator of dimension $3$ as can be seen from its behavior or its
mass square. This is different from the non-supersymmetric flow in
pure $F(4)$ gauged supergravity discussed below in which the
operator dual to the dilaton does acquire an anomalous dimension to
become irrelevant in the IR. It should also be noted that near the
critical points $a$ and $b$ are constant to first approximation.
Therefore, we can use $r\sim \tilde{r}\rightarrow \pm \infty$ near
the critical points.
\\
\indent Note that since the scalar $b(r)$ from the matter multiplets
has $m^2L^2=-6$, the dual operator is given by a scalar mass term.
On the other hand, if the flow had involved the
$(\mathbf{1},\mathbf{3})$ scalars with $m^2L^2=-4$ or $\Delta=4$,
the dual operator would be a fermionic mass term. As a final remark, the central charge
of the five-dimensional conformal field theory is related to the
$AdS_6$ radius by
\begin{equation}
c\sim L^4\sim \frac{1}{V_0^2}\, .
\end{equation}
The ratio of the central charges along this flow is then given by
\begin{equation}
\frac{c_{\textrm{UV}}}{c_{\textrm{IR}}}=\left(\frac{V_0^{(\textrm{IR})}}{V_0^{(\textrm{UV})}}\right)^2
=\left(\frac{g_2^2}{g_2^2-9m^2}\right)^{\frac{3}{2}}
\end{equation}
for $g_1=3m$. This is greater than one for $g_2<-g_1$ with $g_1>0$ or $g_2<g_1$ with $g_1<0$ as
required by the reality condition on $a$ and $b$ at the IR point.
This is in agreement with the c-theorem which we expect to be true
as a general result in holographic RG flows involving only scalars
and the metric \cite{fgpw}.

\subsection{A non-supersymmetric RG flow}
\begin{figure}[!h] \centering
  \includegraphics[width=0.5\textwidth, bb = 0 0 200 150 ]{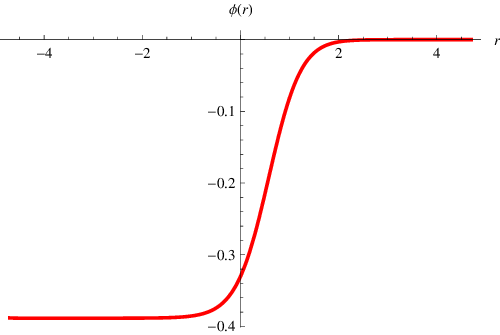}\\
  \caption{$\phi(r)$ solution in pure $F(4)$ gauged supergravity.}
  \label{fig1}
\end{figure}

We now consider a non-supersymmetric RG flow in the pure $F(4)$
gauged supergravity in which there is only one scalar field given by
the dilaton $\sigma$. The dual CFT at the maximally supersymmetric
critical point is given by the 5-dimensional CFT without any global
symmetry corresponding to $E_0$ case of \cite{Seiberg_5Dfield3}.
\\
\indent We will work with the canonically normalized scalar
$\phi=\sqrt{2}\sigma$. In this case,
$L^\Lambda_{\phantom{sa}\Sigma}=\delta^\Lambda_{\phantom{sa}\Sigma}$,
and $g_2$ disappears from the theory. The resulting scalar potential
reduces to that of \cite{F4_Romans}
\begin{equation}
V=m^2e^{-3\sqrt{2}\phi}-4gme^{-\sqrt{2}\phi}-g^2e^{\sqrt{2}\phi}\label{pure_SUGRA_potential}
\end{equation}
where we have denoted the coupling $g_1$ by $g$. There are two
critical points identified in \cite{F4_Romans} long time ago. At the
maximally supersymmetric critical point, we can take $\phi=0$, and
the supersymmetric $AdS_6$ background requires $g=3m$. Using
\eqref{delta_psi} and \eqref{delta_chi}, we can easily see that the
choice $\phi=0$ and $g=3m$ gives $\delta\psi^A_\mu=0$ and
$\delta\chi^A=0$ provided that $\epsilon_{A}$ are Killing spinors on
$AdS_6$. In our convention, at this point, the cosmological constant
is given by $V_0=V(\phi=0)=-20m^2$. The AdS radius is given by
$L=\sqrt{-\frac{5}{V_0}}=\frac{1}{2m}$. Without loss of generality,
we can take $m>0$. The non-trivial critical point is given by
$\phi=-\frac{1}{2\sqrt{2}}\ln 3$, $V_0=-12\sqrt{3}m^2$ and
$L=\frac{\sqrt{5}}{2(3^{\frac{3}{4}})m}$.
\\
\indent We will study an RG flow solution interpolating between
these two critical points with the first and second critical points
identified as the UV and IR fixed points, respectively. The scalar
mass at each critical point can be computed by expanding the
potential to quadratic order in scalar fluctuations. At $\phi=0$, we
find $m_\phi^2L^2_{\textrm{UV}}=-6$ corresponding to the dual
operator of conformal dimension $\Delta=3$. Although the non-trivial critical point
is non-supersymmetric, it is stable as has been shown in
\cite{F4_Romans}. Indeed we can also explicitly compute the scalar
mass at this critical point and find $m^2_\phi L^2_{\textrm{IR}}=10$
which is clearly above the BF bound $-\frac{25}{4}$.
\\
\indent We proceed as in the previous case by taking the metric to
be the domain wall \eqref{DW_ansatz} and the scalar $\phi=\phi(r)$
as a function of the radial coordinate $r$. Since in this case the
flow involves a non-supersymmetric critical point, it is a
non-supersymmetric flow and describes a non-supersymmetric
deformation of the UV SCFT. We cannot use BPS equations to find the
solution. Rather, we need to solve the second order equations of
motion given by varying the action
\begin{equation}
S=\int d^6x
\sqrt{-g}\left(\frac{1}{4}R-\frac{1}{2}\pd_\mu\phi\pd^\mu\phi-V\right)\label{action}
\end{equation}
where the scalar potential $V$ is given by
\eqref{pure_SUGRA_potential}. The field equations arising from the
above action can be found for example in \cite{Skenderis_RG}. In any
case, it is not difficult to derive all equations of motion from
\eqref{action}. With the above ansatz, the equations of motion
become
\begin{eqnarray}
\phi''+5A'\phi'-\frac{dV}{d\phi}&=&0,\label{EOM1}\\
10A'^2-\phi'^2+2V&=&0,\label{EOM2}\\
4A''+10A'^2+\phi'^2+2V&=&0\, .\label{EOM3}
\end{eqnarray}
As in the previous case, $A'=\frac{dA}{dr}$ and
$\phi'=\frac{d\phi}{dr}$ ect. It is well-known that only two of them
are independent. We will use the first two equations to solve for
the solution. Note also that these equations also imply
\begin{equation}
A''=-\frac{1}{2}\phi'^2\, .
\end{equation}
\indent It is not known how to obtain analytic solutions to these
equations. Therefore, as in \cite{F4_nunezAdS6}, a numerical
solution will be given here. We start by setting $g=3m$. The
boundary conditions at the UV and IR points for the scalar field are
$\phi_{\textrm{UV}}=0$ and $\phi_{\textrm{IR}}=-\frac{1}{2\sqrt
2}\ln 3=-0.3884$. The equations to be solved are given by
\begin{eqnarray}
\phi''(r)+5 \phi'(r) A'(r)+3 \sqrt{2} m^2 e^{-3 \sqrt{2} \phi(r)}-12
\sqrt{2} m^2 e^{-\sqrt{2} \phi(r)}+9 \sqrt{2} m^2 e^{\sqrt{2}
\phi(r)}&=&0,\,\,\,\,\,\,\,\,\,\,\,\,\\
10 A'(r)^2-\phi'(r)^2-2 m^2 e^{-3 \sqrt{2} \phi(r)} \left(12 e^{2
\sqrt{2} \phi(r)}+9 e^{4 \sqrt{2} \phi(r)}-1\right)&=&0\, .\,\,\,\,
\end{eqnarray}
\indent Without loss of generality, we will choose $m=1$ in the
solution given below since in the above equations, $m$ can be
absorbed into the radial coordinate $r$ by scaling $r\rightarrow
mr$. By using appropriate boundary conditions and a computer program
\textsl{Mathematica}, we can find a solution for $\phi$ as shown in
Figure \ref{fig1}. This is the same as the solution obtained in
\cite{F4_nunezAdS6}, and it is clearly seen that the solution
interpolates between the two critical points.
\\
\indent We now look at the asymptotic behavior of the solution near
the critical points. At the UV point, we take $r\rightarrow \infty$
and $A(r)\rightarrow \frac{r}{L_{\textrm{UV}}}$ with
$L_{\textrm{UV}}=\frac{1}{2m}$. The linearized equations of motion
give
\begin{equation}
\phi\sim
c_1e^{-\frac{3r}{L_{\textrm{UV}}}}+c_2e^{-\frac{2r}{L_{\textrm{UV}}}}\,
.\label{asym1}
\end{equation}
In this case, we find again that the dual operator should be given by a mass term of the scalars in hypermultiplets.
\\
\indent The solution with $c_1=0$ and $c_2=0$ corresponds to non-normalizable and normalizable modes,
respectively. The flow with $c_1=0$ is driven by turning on a mass
term for hypermultiplet scalars while the flow with $c_2=0$ is
driven by a vev of this operator. To decide between these two
possibilities, we generally need to compute the ``superpotential'' in the
Hamilton-Jacobi formalism for holographic RG flows
\cite{deBoerHamilton1}, \cite{deBoerHamilton2}.
\\
\indent However, in this case, there is a simple argument in
deciding between a vev deformation and a true deformation by turning
on a dual operator as discussed in \cite{F4_nunezAdS6}. The scalar
field is negative ($\phi<0$) along the flow. There cannot be any vev
deformations since the physical vacuum expectation value,
$\langle\mc{O}_\phi\rangle$, of a positive definite operator must be
positive. Therefore, the scalar $\sigma$ corresponds to turning on a
relevant operator of dimension $\Delta=3$ in the dual UV CFT. This
is described by the non-normalizable mode
$e^{-\frac{2r}{L_{\textrm{UV}}}}$ in \eqref{asym1}.
\\
\indent The situation is very similar to the RG flow in minimal gauged supergravity in seven
dimensions studied in \cite{non_SUSY7Dflow}. The similarity can be
made even more explicit by finding the behavior of $\phi(r)$ near the
UV point using the ``true'' superpotential. This gives $\phi(r)\sim
e^{-\frac{3r}{L_{\textrm{UV}}}}$ which precisely corresponds to the
vev of a dimension $3$ operator.
\\
\indent At the IR point, we find
\begin{equation}
\phi\sim e^{\frac{\sqrt{65}+5}{2}\frac{r}{L_{\textrm{IR}}}}
\end{equation}
with $L_{\textrm{IR}}=\frac{\sqrt{5}}{2m (3^{\frac{3}{4}})}$. We see
that the corresponding operator is irrelevant and has dimension
$\Delta=\frac{\sqrt{65}+5}{2}>5$. As pointed out in
\cite{F4_nunezAdS6}, the dual operator acquires an anomalous
dimension in the IR. The ratio of the central charges in this case
is given by
\begin{equation}
\frac{c_{\textrm{UV}}}{c_{\textrm{IR}}}=\frac{27}{25}\, .
\end{equation}
\section{Conclusions}\label{conclusion}
In this paper, we have studied the scalar potential of $F(4)$ gauged
supergravity coupled to three vector multiplets and the
corresponding critical points. We have found new critical points
apart from the known trivial one with maximal supersymmetry. One of
these critical points is supersymmetric with $SO(3)$ symmetry and
can be considered as a gravity dual to a new supersymmetric
SCFT$_5$. This should be an IR fixed point of the $N=2$ CFT$_5$ with
symmetry $F(4)\times SO(3)$. We have also given an analytic solution
describing the RG flow between these two critical points. This flow
involves two active scalars corresponding to relevant operators of
dimension $3$ in the dual field theory. The two scalars transform
under $SU(2)_R\times SO(3)$ as $(\mathbf{1},\mathbf{1})$ and a
combination of $(\mathbf{3},\mathbf{3})$ representations. We have
identified these operators with hyper-scalar mass terms. We have
shown that the flow is driven by a vacuum expectation value of these
operators. Furthermore, we have shown that the non-supersymmetric
critical point of the pure $F(4)$ gauged supergravity is unstable in
the matter coupled theory.
\\
\indent It is interesting to study the matter coupled $F(4)$ gauged
supergravity with other global symmetry group $E_{N_f+1}$ in more
details. This could give some insights to the dynamics of the
D4-brane worldvolume theory in the presence of D8-branes.
Furthermore, it would be useful to uplift the flow solution to ten
dimensions in the context of massive type IIA theory in which the
$F(4)$ gauged supergravity can be embedded
\cite{Pope_6D_massiveIIA}. The solution should also be uplifted to
type II theory by the truncation given in \cite{Eoin}. With an
analytic solution given here, the uplifted solution would be useful
in doing a brane probe computation which could give some insight to
D-brane dynamics and the corresponding gauge theory.
\\
\indent It is remarkable that we have found a supersymmetric $AdS_6$
critical points although a recent result in \cite{AdS6_from10D}
has pointed out the scarcity of supersymmetric $AdS_6$ solutions
from massive type IIA supergravity. This makes the embedding of the
new supersymmetric $AdS_6$ solution discovered here in massive type
IIA (if possible) highly non-trivial. Along the line of finding
supersymmetric $AdS_6$ solutions, we should study the matter
coupled $F(4)$ gauged supergravity in more details with possibly
different gauge groups to see whether there are any possibilities of
$AdS_6$ critical points. Moreover, there is another possibility to
obtain supersymmetric $AdS_6$ by using Hopf T-duality, see for
example a discussion in \cite{Pope_Hopf}, in type IIB theory
\cite{Eoin_private}.

\acknowledgments I gratefully thank Henning Samtleben for helpful
correspondences and valuable comments. I would also like to thank
Eoin O. Colgain for pointing out reference \cite{F4_nunezAdS6} and a
number of useful discussions. Invaluable suggestions from Carlos
Nunez are gratefully acknowledged. This work is partially supported
by Thailand Center of Excellence in Physics through the
ThEP/CU/2-RE3/11 project and Chulalongkorn University through
Ratchadapisek Sompote Endowment Fund.

\end{document}